# Accelerating Statewide Connected Vehicles Big (Sensor Fusion) Data ETL Pipelines on GPUs

**Abdul Rashid Mussah**
Department of Civil and Environmental Engineering
University of Missouri, Columbia, MO 65211
Email: akm2fx@mail.missouri.edu

**Maged Shoman**
Department of Civil and Environmental Engineering
University of Missouri, Columbia, MO 65211
Email: mas5nh@mail.missouri.edu

**Mark Amo-Boateng**
Department of Civil and Environmental Engineering
University of Missouri, Columbia, MO 65211
Email: marbz@missouri.edu

**Yaw Adu-Gyamfi**
Department of Civil and Environmental Engineering
University of Missouri, Columbia, MO 65211
Email: adugyamfiy @missouri.edu

Word Count: 4,600 words + 3 table (250 words per table) = 5,350 words

*Submitted May 8, 2023*

**ABSTRACT**
Real-time traffic and sensor data from connected vehicles have the potential to provide insights that will lead to the immediate benefit of efficient management of the transportation infrastructure and related adjacent services. However, the growth of electric vehicles (EVs) and connected vehicles (CVs) has generated an abundance of CV data and sensor data that has put a strain on the processing capabilities of existing data center infrastructure. As a result, the benefits are either delayed or not fully realized. To address this issue, we propose a solution for processing state-wide CV traffic and sensor data on GPUs that provides real-time micro-scale insights in both temporal and spatial dimensions. This is achieved through the use of the Nvidia Rapids framework and the Dask parallel cluster in Python. Our findings demonstrate a 70x acceleration in the extraction, transformation, and loading (ETL) of CV data for the State of Missouri for a full day of all unique CV journeys, reducing the processing time from approximately 48 hours to just 25 minutes. Given that these results are for thousands of CVs and several thousands of individual journeys with sub-second sensor data, implies that we can model and obtain actionable insights for the management of the transportation infrastructure.

**INTRODUCTION**

Innovation in managing transportation infrastructure at local and state levels is widely believed to be the next breakthrough needed to support the growing population demands on the already limited available infrastructure [1]. Currently, the information needed for traffic modelling, monitoring and prediction is sourced from legacy infrastructure and equipment that are not scalable and reliable as the demands for real-time insights grow [2], [3]. Connected vehicles (CV) promise to be a technology that will provide live fingerprinting and activity monitoring on all transportation infrastructure and allow state agencies known to make insights into the happenings on the road. These data from CV have to potential to transform traffic modelling and prediction for efficient infrastructure management and will lead to benefits including congestion management, improved signal intersection management, increased traffic throughput, advanced traffic predictions, reduced traffic incidents, and faster response times and alternate routing of traffic to uneventful traffic events on the roads [2]–[6]. The challenge, however, is that even at the current stage where CVs are in the minority big data from CVs and related sensors are so much that their processing and analytics take several hours to days, and we may not realize the intended benefits on our road infrastructure.

The management of transportation infrastructure is undergoing a major shift as the future moves towards utilizing real-time big data streaming from sensors embedded into connected vehicles (CV). Vehicle manufacturers are rapidly incorporating CV technology into new and existing vehicles for several advantages, including automation and navigation, as well as kinematic sensor and driver behavior monitoring, live over-the-air updates through wireless communications, advanced road warnings of traffic flow conditions [7], as well as improved fuel and battery efficiency [2], [5]. Governments and other state and local institutions such as the departments of transportation (DOTs) responsible for creating, maintaining, and managing road infrastructure can benefit from the abundance of CV data to gain insight into real-time instances of situations occurring on the roads and make informed decisions regarding response to traffic flow conditions and the situation of roadway infrastructure. However, the challenge remains on how to effectively process the very large volumes of this CV data at a statewide level to capitalize on the potential of deploying an efficient and continuously updateable transportation infrastructure management framework.

Transportation infrastructure, including road pavement surface data, road networks, signals and intersections, as well as parking, has been an indicator and driver of the modern urban economic growth in cities. Until recently, and in compliance with Moore's Law that promises the doubling of computing power roughly every 18 months [8], processing and analyzing data from transportation infrastructure had not been a challenge. CPUs have been developed and have been at the forefront of data processing. However, Moore's Law is believed to have reached its physical limits, and the introduction and rise in big data across several industry verticals such as finance, social networks, transportation, retail, telecommunications, and biology have meant that new and innovative approaches be adopted to process and analyze these data for actionable insights. Recent research has been focused on finding new and alternative methods for crunching big data for actionable insights [9]–[11]. These approaches include parallel computing, edge computing, data processing on graphical processing units (GPU), and quantum computing (still in the infancy of its development). Currently, processing big CV and sensor data on traditional computing systems take several hours to days, and we wonder if maturing alternative technologies like GPU could deliver better performance for CV big data of transportation infrastructure at the county, district and state levels [10], [11].

Graphics Processing Units (GPU), originally designed to accelerate graphics calculations on computers, is a massively parallel computing device. The introduction of Nvidia Compute Unified Device Architecture (CUDA) ushered in the era of massive parallel scientific computations on GPUs originally intended for gaming and visual graphics processing [12]–[14]. Although GPUs are limited in their computing capabilities, the massively parallel architecture offers significant speedups for simple data processing tasks with independent execution paths, also known as Single Instruction Multiple Data (SIMD) (see Figure 1). Exploiting the massive parallelism of GPUs usually means launching millions of threads on thousands of processing cores on a single GPU [15], [16]. The bottlenecks in using GPUs for data processing usually include limited GPU memory and data transfer bandwidth between the CPU and GPU. There are however tricks to overcome or hide these challenges inherent in GPUs like simultaneous data transfer and processing of batched data. Despite these challenges with GPUs, they are known to accelerate scientific computations up to 200,000x over CPUs when leveraged correctly [12], [14], [17]. As such, GPUs and related accelerator hardware technologies now form the backbone of all supercomputing infrastructure.

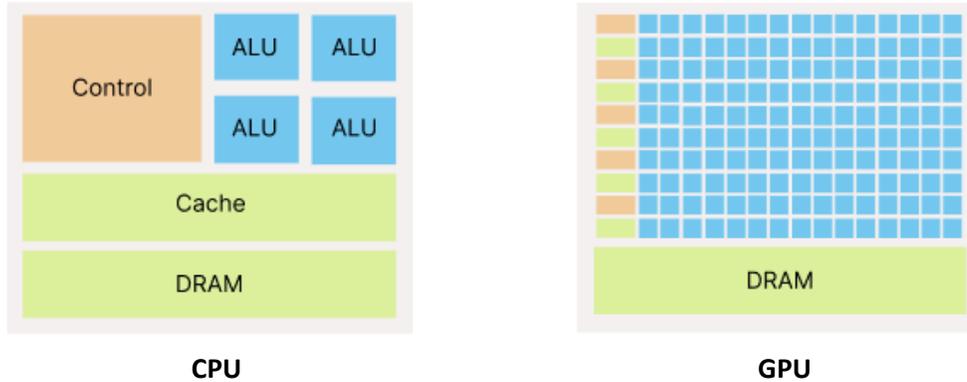

**Figure 1 CPU and GPU architectures**

The traditional big data engineering pipeline includes: extract – data extraction from multiple data sources; transform – feature engineering and data transformation of the extracted data; load – data formatting and presentation to analytics and engineering models [11]. This process known as ETL (extract, transform and load) is a standard data processing pipeline in big data applications. In this paper, we present a framework (an algorithm) to accelerate the traditional data engineering pipelines: extract, transform, and load (ETL) as applied to big CV data at local and state levels on GPUs that will lead to real-time actionable insights on state-wide transportation infrastructure.

**METHODS**
This section covers the background and methods used in the design of the framework for accelerated big CV data in this paper, including the programming language frameworks, dataset description, and the new algorithm for accelerated ETL of big CV data.

**Connected Vehicle Dataset**
Data for this research was provided by the Missouri Department of Transport research project on Connected Vehicles. The project collected CV data from 100,000 voluntary participants during the month of May 2021. Each CV data includes a unique journey identifier and has a time resolution of 0.05 seconds. Other sensor data collected included GPS data: longitude, latitude, altitude, speed, heading (direction), IMU sensor data, acceleration, zip code, and other vibration sensor data. Streamed CV data for each of the vehicles were stored in separate files on Amazon S3 in text format (CSV) and averaged about 2,000 text files per day – occupying a minimum of 100GB of text day each day. The total data size for the study period was over 50 TB with over 1,000,000 text files. Other data that was used in this research include the Waze incident report data (WIR) and the Police crash report (PCR) data. The data covered the entire road network of the state of Missouri including major roads and arterial roads. Aside from merging and querying different file data structures, one of the key challenges with the processing of this CV data is the tracking of unique CV data across several thousands of files in different folders and different road networks in different zip codes. A snapshot of the data from one file is shown in Table 1 below.

**TABLE 1 Snapshot of CV Data used in this research.**

| Journey Id | Timestamp | Latitude | Longitude | Postal Code | Speed | Heading |
|---|---|---|---|---|---|---|
| 33456rd | 2021-05-09 03:48:42 | 37.664087 | -92.6546 | 65536 | 105.98 | 33 |
| 31224tf | 2021-05-09 03:49:42 | 37.667707 | -92.6490 | 65536 | 0 | 53 |
| 22124fs | 2021-05-09 03:49:49 | 37.690978 | -92.6490 | 65536 | 48.38 | 33 |
| 33456rd | 2021-05-09 03:48:42 | 37.664087 | -92.6546 | 65536 | 105.98 | 33 |

**Nvidia Rapids and Dask Framework**

Nvidia Rapids [18], [19] is an open-source software library that allows developers to build and run data science and analytics applications on top of GPUs (graphics processing units). It is designed to make it easier for developers to leverage the power of GPUs to accelerate the performance of their applications, particularly when working with large datasets.

Nvidia Rapids includes a variety of tools and libraries for tasks such as data processing, machine learning, and data visualization. It is designed to work with a range of programming languages, including Python, R, and C++, and can be integrated with other software libraries and frameworks such as Dask.

Dask [18], [19] is a flexible parallel computing library for analytics in Python. It allows users to harness the full power of their CPU and memory resources without the need to rewrite code or switch to a distributed system. Dask provides a simple and powerful way to parallelize existing Python code and scale it up to handle large datasets and computations. It can be used to build custom parallel algorithms and integrate with other libraries such as NumPy, Pandas, and Scikit-learn. Dask can be used on its own or in combination with other distributed computing tools such as Apache Spark.

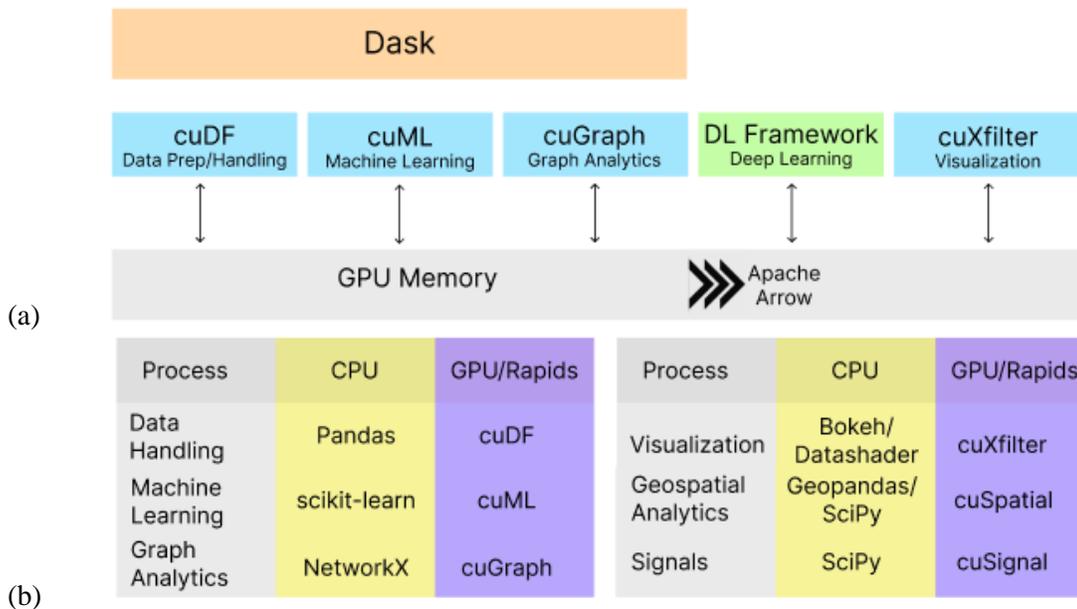

(a)

(b)

**Figure 2 Nvidia Rapids framework. (a) Dask interaction with Rapids (b) Nvidia Rapids libraries in comparison to popular corresponding python data science libraries**

**Accelerated ETL Algorithm/Framework for Spatio-Temporal CV Data**

Figure 3 below showcases the framework utilized to accelerate the large CV data presented in the paper. The process for fusing CV and sensor data is outlined in the accompanying diagrams. Our approach involved collecting data from multiple CV sources across several folders and drives, temporarily storing it in a massive in-memory database, transforming them into a multidimensional spatio-temporal lattice with distinct attributes for attribute-based hyper-dimensional analysis. The process employed to achieve this capitalized on the Dask framework for efficient processing and filtering of embarrassingly large data, utilizing a GPU backend powered by Nvidia Rapids. To begin, a local cluster was established and the Dask framework was utilized to load all sensor data files collected from CVs for a particular period (1). The data was then filtered to include only the columns of interest and collected into a massive in-memory data lake (2). At this point, a unique index was calculated for the data which was later re-partitioned to optimize performance and cut down on the number of deployed Dask workers (2).

To begin the process of translating the data into a multidimensional spatio-temporal matrix, unique indices for each data row were computed as presented in the procedure outlined in (3). This involved computing unique spatial discretized bins for longitude and latitude, and using each data row's longitude and latitude to calculate its spatial positional index, which was then placed into the appropriate bin. A similar process was used to create a discretized temporal bin based on the day, hour, and minute. With these unique spatial, temporal, and directional indices, unique unrolled positional global indices were calculated for each data point, which was then used to translate the in-memory database into the 3D spatial-time lattice (4).

Each cube in the 3D spatial-time lattice (5) contained all data entries with the same index, as well as attributes such as speed and direction, which could then be used for hyper-dimensional data operations based on the data's attributes (6). The study utilized filtered and stacked data based on these attributes, and performed further analytics based on the attributes of speed, data counts, and direction.

**Experimental Setup for Spatio-Temporal CV Data Processing**

The experimental setup for this project was conducted on AWS GPU virtual machines equipped with Intel Xeon Platinum 8259CL 48-core vCPUs at 2.50GHz, 192 GB of RAM, and 4 T4 GPUs with 16 GB VRAM each. The virtual machines were running the AWS optimized Ubuntu 18.04 LTS operating system and had the CUDA 10.2 software with a driver version of 440.33.01 installed. Docker CE v18.03.1-ce and Nvidia Docker2 were also included in the software stack for GPU containerization. The DLI RAPID Course - Base Environment container image v1.0.0 from the Nvidia Container Catalogue (NGC) was utilized to launch a Python Jupyter Lab environment on the AWS virtual machine. This container image came pre-installed with software such as Rapids, Conda, Graphviz, cuDF, cuPy, and more, streamlining the experimental setup. The pull and launch of the container image also provided internal port access to the container and allowed for global internet access to the Jupyter Lab environment outside the localhost.

To evaluate the performance of the experimental setup, the built-in Python timeit() function with the repeat() method was used to run each algorithm multiple times. The average, best, and worst-case run times along with their standard deviations were recorded and analyzed. The results of these experiments, including various optimizations for accelerating ETL pipelines for big CV data, are presented in Table 2.

**TABLE 2 Experimental setup used in this research**

| # | CV Data | Name | Platform |
|---|---|---|---|
| 1 | Speed | Data Binning | CPU/GPU |
| 2 | | Indexing - Latitude | CPU/GPU |
| 3 | | Indexing - Longitude | CPU/GPU |
| 4 | | Normalization | CPU/GPU |
| 5 | | Data Export | CPU/GPU |
| 6 | Volume | Reduction - Count | CPU/GPU |
| 7 | | Reduction – Sum | CPU/GPU |
| 8 | | Indexing – Latitude | CPU/GPU |
| 9 | | Indexing – Longitude | CPU/GPU |
| 10 | | Filter | CPU/GPU |
| 11 | | Normalization | CPU/GPU |
| 12 | | Data Export | CPU/GPU |

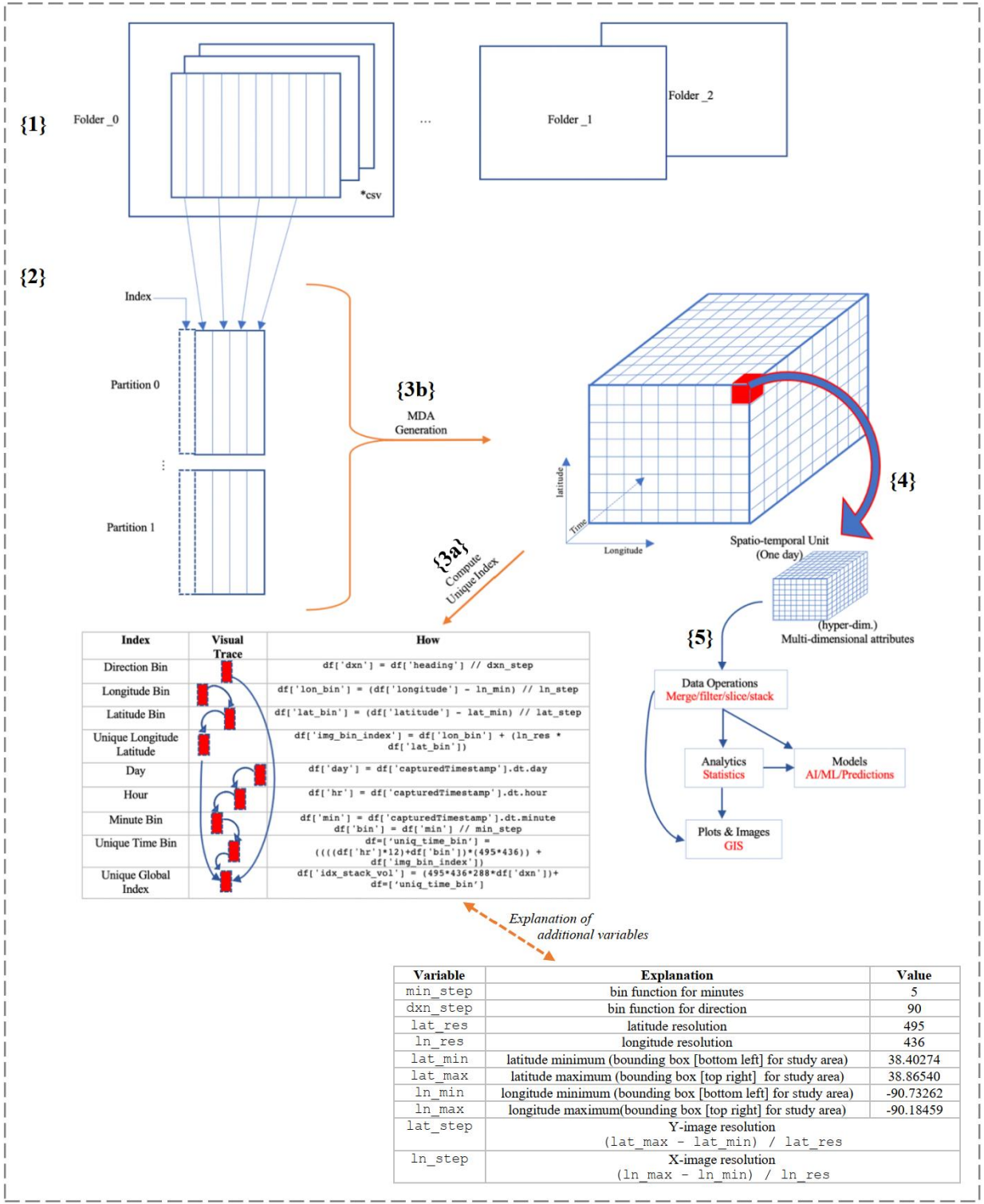

**Figure 3 ETL for Connected Vehicle Data**

**RESULTS**

This section presents the performance evaluation of accelerating the ETL pipeline of big CV data under various algorithm optimizations as discussed in the preceding section. The section presents the running times and the speedups when running the experiments under various scenarios.

**Comparison of CPU versus RAPIDs GPU Source Code**

The original code for the preprocessing of the big CV data is shown in Figure 4 below. Here, the code initializes standard libraries and then proceeds to implement the data preprocessing logic for traffic speed and volume as shown in Table 2. Figure 4 shows the pseudocode for processing the data binning of spatial traffic data to a 2D image array utilizing CPU enabled python libraries. Similarly, figure 5 shows achieving the same results using the RAPIDs framework. After the standard libraries have been initialized and set up on a GPU cluster, figure 5 shows the minimal process required for binning the spatial traffic data to a 2D image array when deployed with the GPU framework. The RAPIDs GPU code shows a simpler programming API when compared to the native CPU code.

---

**Algorithm 1:** Sample Code from CPU Pipeline for processing CV big data

1. **Procedure:** BINNING SPATIAL POINTS
2. $d \leftarrow$ (file data eg. '2021-02-08')
3. $h \leftarrow$ (hour of day, range (0-23))
4. $t \leftarrow$ range (1, 13)
5. loop: (t) in range
6. $df \leftarrow 'timebins'_{csv} + '/' + 'str(n)' + '/' + 'str(d)' + '/' + 'str(h)' + '/' + str(t)'$
7. $df \leftarrow df[['latitude','longitude','speed']]$
8. $df['latitude'] \leftarrow df['latitude'] * -1$
9. $x_{cut} \leftarrow pd.cut\ (df.latitude.np.linespace\ (min(df.latitude), max(df.latitude)), right = False$
10. $y_{cut} \leftarrow pd.cut(df.longitude.np.linespace(min(df.longitude), max(df.longitude)), right = False$
11. $df \leftarrow df.groupby([x_{cut}, y_{cut}]).mean()$

---

**Figure 4 Sample source pseudocode for big CV data on CPUs: Involving spatial binning of traffic geocoded data**

---

**Algorithm 2:** Sample Code from Dask + Rapids GPU Pipeline for processing CV big data

1. **Procedure**: BINNING SPATIAL POINTS
2. $df['bin'] \leftarrow df['min']//min_{step}$
3. $df['dxn'] \leftarrow df['heading']//dxn_{step}$
4. $df['lat_{bin}'] \leftarrow (df['latitude'] - lat_{min})//lat_{step}$
5. $df['lon_{bin}'] \leftarrow (df['longitude'] - lon_{min})//lon_{step}$

---

**Figure 5 Sample source pseudocode for big CV data on GPUs using RAPIDs and Dask CUDA: Involves data binning of traffic speed data on GPU, as well as indexing of traffic data along latitude and longitude on GPU.**

**Performance Evaluation of the Running Times**

This section presents the performance evaluation of accelerating the ETL pipeline of big CV data under various algorithm optimizations as discussed in the preceding section. The section presents the running times when running the experiments under various scenarios as shown in Table 2. Here we present the summaries of 25 runs for each experiment on CPUs and GPUs (see Table 3). In general, there is a marked speed improvement from 25.6 times to 72.2 times faster when the experiment is carried out on the GPU. Overall, the average experiment lasted close to 42 hours on the CPU while the GPU lasted 25 minutes.

**TABLE 3 Running times of the ETL algorithms by number of CV data in seconds**

|  | Metric | Metric | CPU (seconds) | GPU (seconds) | Speedup (X) |
|---|---|---|---|---|---|
| Data Binning (Bucketing) | Speed | Avg | 31207.3037 | 442.9241631 | 70.457 |
|  |  | Min | 28250.79274 | 399.2104539 |  |
|  |  | Max | 34100.80986 | 482.9444833 |  |
|  |  | Std. Dev. | 1610.113766 | 28.54695702 |  |
|  | Volume | Avg | 270.3653968 | 4.525327381 | 59.745 |
|  |  | Min | 244.8511661 | 3.759135161 |  |
|  |  | Max | 292.5079318 | 4.993279367 |  |
|  |  | Std. Dev. | 14.57223785 | 0.303196382 |  |
| Data Indexing | Latitude (Speed) | Avg | 17302.22604 | 248.4833932 | 69.631 |
|  |  | Min | 15621.62386 | 222.116116 |  |
|  |  | Max | 19099.29555 | 268.877181 |  |
|  |  | Std. Dev. | 978.0325212 | 14.20678187 |  |
|  | Latitude (Volume) | Avg | 20386.04865 | 295.9992489 | 68.872 |
|  |  | Min | 18430.08283 | 269.4204727 |  |
|  |  | Max | 22027.04409 | 314.4327911 |  |
|  |  | Std. Dev. | 1103.700135 | 13.92043728 |  |
|  | Longitude (Speed) | Avg | 16096.90286 | 222.9147291 | 72.211 |
|  |  | Min | 14394.11849 | 202.2305935 |  |
|  |  | Max | 17544.64725 | 246.6807316 |  |
|  |  | Std. Dev. | 971.4204992 | 14.31896136 |  |
|  | Longitude (Volume) | Avg | 18564.8494 | 260.4397439 | 71.283 |
|  |  | Min | 17018.96634 | 236.0678421 |  |
|  |  | Max | 20318.35802 | 285.409977 |  |
|  |  | Std. Dev. | 987.9884258 | 15.64979372 |  |
| Normalize | Speed | Avg | 1148.602237 | 16.63911158 | 69.030 |
|  |  | Min | 1048.054805 | 15.35796017 |  |
|  |  | Max | 1245.09652 | 17.95620487 |  |
|  |  | Std. Dev. | 63.48909217 | 0.862589958 |  |
|  | Volume | Avg | 704.0662949 | 10.10884161 | 69.649 |
|  |  | Min | 640.6291442 | 9.028229595 |  |
|  |  | Max | 765.6214614 | 10.99522392 |  |
|  |  | Std. Dev. | 38.22872458 | 0.583591657 |  |
| Reduction | Count Unique (Volume) | Avg | 37.46484825 | 1.465020072 | 25.572 |
|  |  | Min | 34.21559376 | 0.531817414 |  |
|  |  | Max | 41.60869151 | 1.979728872 |  |
|  |  | Std. Dev. | 2.099328467 | 0.360129996 |  |
|  | Sum (Volume) | Avg | 42178.2191 | 599.0471511 | 70.408 |
|  |  | Min | 37627.96991 | 535.4793912 |  |
|  |  | Max | 46179.47844 | 647.9161687 |  |
|  |  | Std. Dev. | 2641.918263 | 31.79059656 |  |
|  | Filtering (Volume) | Avg. | 826.4330289 | 11.8381508 | 69.811 |
|  |  | Min | 752.7677184 | 11.03521217 |  |
|  |  | Max | 886.7951491 | 12.98443984 |  |
|  |  | Std. Dev. | 38.37536422 | 0.541540024 |  |
| Data Export | Speed | Avg | 250.7947809 | 4.403811729 | 56.949 |
|  |  | Min | 232.204561 | 3.53902743 |  |
|  |  | Max | 276.381477 | 4.991058649 |  |
|  |  | Std. Dev. | 13.55633277 | 0.343096092 |  |
|  | Volume | Avg | 142.4564327 | 2.553969987 | 55.778 |
|  |  | Min | 124.8083052 | 1.951443981 |  |
|  |  | Max | 152.3210562 | 2.971581714 |  |
|  |  | Std. Dev. | 7.463689586 | 0.265524497 |  |
| Overall |  | Avg | 149115.733 | 2121.342 | 70.293 |

**Transformed Data Output**

A multidimensional data output for a temporal analysis unit can be visualized as an image. In the experiment carried out, 5-minute time bins were employed for the data binning. Speeds and volumes were computed for the batched data in the four cardinalities of North, East, West and South. The process resulted in 8 channels in total for each 5 minute batch. These channels were then composited into a single image for visualization purposes. The data in its channelized form was then stored for analysis purposes in a hierarchical (hdf5) data format. The process allowed for data compression of over 50 TB of one year's worth of CV data, into less than 20 GB of data in the multidimensional spatio-temporal format. The final output data maintains a resolution and structure suitable for implementation in many deep learning involved modeling processes. CNNs, ConvLSTMs and other encoder-decoder deep architectures like UNets have been employed on the data in this form with high levels of accuracy in the instance of network level traffic forecasting [20], [21]. Figure 6 below shows the individual and combined heading channel outputs for the two traffic flow variables, as well as a final composite visualization of the data as represented in image format.

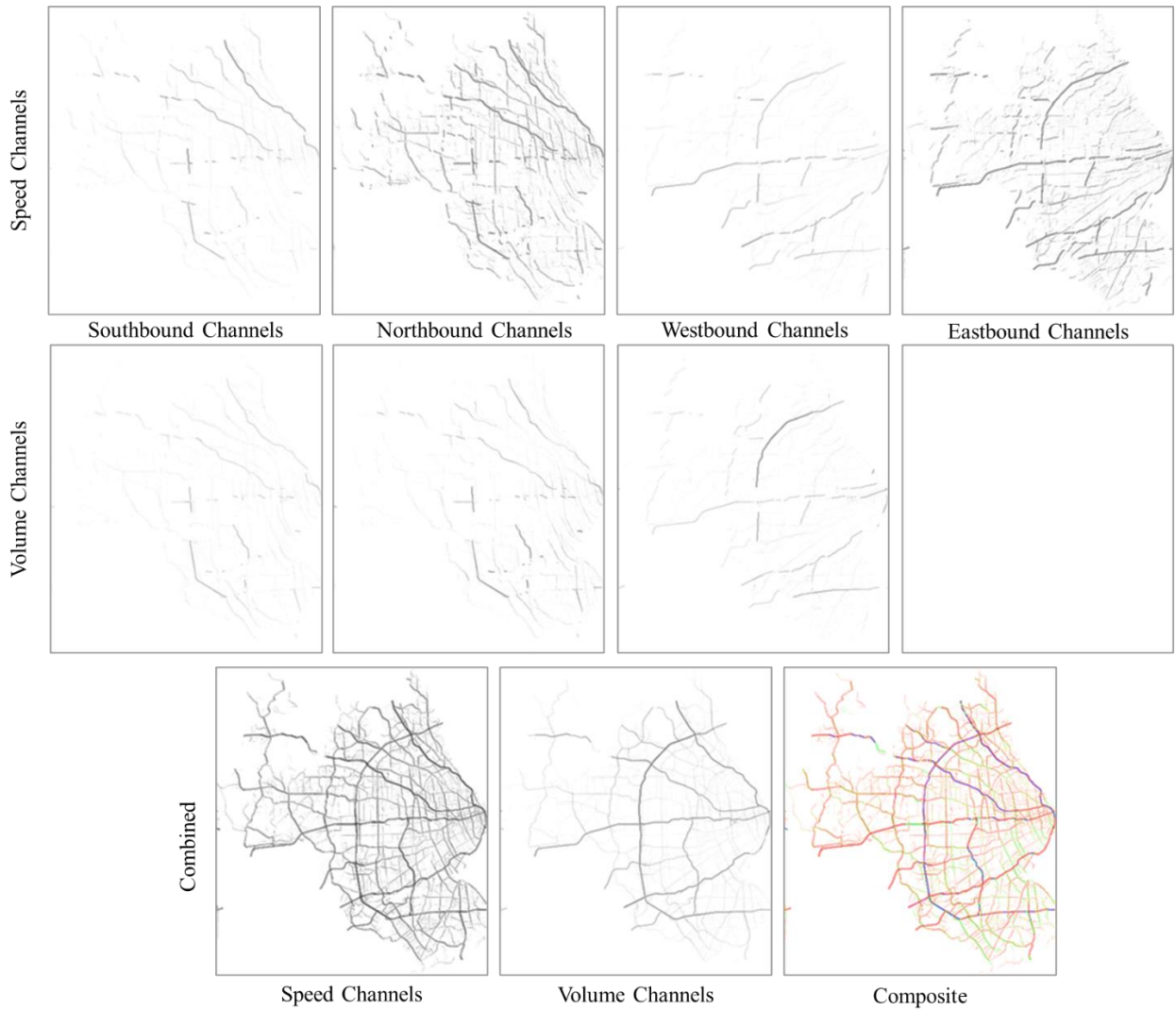

**Figure 6 Sample Output of 5-minute batch processed CV data for Speeds and Volumes, and final composite image visualization.**

**DISCUSSION**

The development and construction of high-performance computing systems and supercomputers have become a priority for most advanced countries [1], [19]. With the sudden rise in data generating devices and the availability of stable interconnected networks, the new age of the networked community is upon us. The rapid adoption of EVs has also ushered in a new age of connectivity for road vehicles. This brings several advantages of having real-time sensor data streams sent to vehicle authorities and security agencies to improve safety and responsiveness to emergencies. However, working with huge data from connected vehicles continue to pose a challenge for everyone. In our experiment, processing data from 1,500 unique journeys by the connected vehicles for a day usually takes about 2 days to be processed. This 48-hour lag time between data sensing and interpretation offers no benefit to the transportation authorities and security agencies. In this light, several attempts are being made to find faster and more efficient ways of working with big data from connected vehicles. The upcoming approaches include the use of quantum computing and GPUs.

For this work, we focused on GPUs as they are matured and are more readily available and cost-effective. In the experiments with the GPUs, we noticed up to 72 times faster execution than compared to traditional CPU-based processing. Our results confirm what other researchers have reported in scientific fields. For instance, one investigator reported up to 400X speedup in using GPUs to accelerate his computational workloads [22], [23]. In a recent flood forecasting study, the authors reported a speed up between 80X and 88X in their work with GPUs. Similarly, works with image processing have shown to have a speedup between 10x and 20x when using GPUs as coprocessors.

Our work, with a speedup of 70x, therefore goes to confirm what has been recently cited by several academic investigators. We, therefore, believe that with the general availability of GPUs becoming mainstream, they can be leveraged to accelerate difficultly and demanding computational tasks in the transportation industry. In this work, we have used the Python-based RAPIDS framework with Dask CUDA to build an end-to-end pipeline and have achieved a modest 70x fold speedup (reducing the execution time from 48 hours to 25 minutes). We believe our research will throw more light on the potential applications of using GPUs in harnessing the insights that connected vehicle big data can afford us.

**CONCLUSIONS**

In this research, we explored accelerating Big CV Data Pipelines on GPUs using the RAPIDS framework and Dask parallel processing, transforming point CV data into multidimensional spatiotemporal array format for implementation in deep neural network architectures. Our preliminary results show a reduction in computational time of processing 24 hours' worth of data into 5 min image stacks, from 41 hours to 35 minutes, accelerating the entire process by 70 times. In addition, aside from the initial library imports and cluster setup codes, the RAPIDS and Dask framework made the source codes relatively simpler, reducing most of the computations to a single line of code. In contrast, the original CPU code took several lines of code (about 20 for each task). In summary, as the number of connected vehicles and sensors increases in our community, the need for real-time sensor data processing and data fusion will continue to remain a challenge. Given this, we believe that leveraging modern tools such as GPUs and other accelerators can provide an inexpensive pathway for processing these data in real-time, which will allow for easy implementation with more advance computational processing frameworks. This will help us gain insights in real-time and also improve the management and safety of our transportation infrastructure.

**AUTHOR CONTRIBUTIONS**
The authors confirm their contribution to the paper as follows: study conception and design: A. R. Mussah, M. Amo-Boateng, Y. Adu-Gyamfi, M. Shoman; data collection: A. R. Mussah, M. Amo-Boateng, Y. Adu-Gyamfi, M. Shoman; analysis and interpretation of results: M. Amo-Boateng, Y. Adu-Gyamfi, A. R. Mussah; draft manuscript preparation: A. R. Mussah, M. Amo-Boateng, Y. Adu-Gyamfi. All authors reviewed the results and approved the final version of the manuscript.